\documentstyle[11pt]{article}
   \newcommand {\nc}{\newcommand}
   \nc{\eq}{\begin{equation}}
   \nc{\en}{\end{equation}}
   \nc{\eqa}{\begin{eqnarray}}
   \nc{\ena}{\end{eqnarray}}
   \nc{\eqann}{\begin{eqnarray*}}
   \nc{\enann}{\end{eqnarray*}}
   \def\prf{{\bf{Proof:}}\\}
   \def\endprf{${\bf{\Box}}$\\}
   \newtheorem{lemma}{Lemma}
   \newtheorem{theorem}{Theorem}

   \nc {\dfn}[1]{{\it{#1}}}
   \nc{\nn}{\nonumber}
   \def\dlt{\delta}
   \def\ep{\epsilon}
   
   \def\etal{{\it et al} }

   \nc{\sqt}{\sqrt{2}}
   \nc{\tsqt}{2\sqt}
   \nc{\msqt}{$\sqt$}
   \nc{\nsqt}{$-\sqt$}
   \nc{\mtsqt}{$\tsqt$}
   \nc{\ntsqt}{$-\tsqt$}

   \def\ot{\otimes}
   
   \def\smdp{>\hspace{-0.2cm}\lhd}

   \nc {\inv}[1]{#1^{-1}}
   \nc {\hc}[1]{{#1}^\dag}
   \nc {\cc}[1]{{#1}^\ast}
   \nc {\ad}[2]{Ad_{#1}({#2})}
   \nc {\wad}{\widetilde{Ad}}
   \nc {\wadf}[2]{\widetilde{Ad}_{#1}({#2})}
   \nc {\pb}[1]{{#1}^\ast}
   \def\tld{\tilde}
   \def\pr{\prime}

   \def\natural{{\cal N}}
   \def\intg{{\cal Z}}
   
   \def\complex{{\cal C}}
   
   \nc {\ga}[2]{{#1}[{#2}]}
   \nc {\cga}[1]{\ga{\complex}{#1}}
   \nc {\cgag}{\cga{G}}
   \nc {\eu}[1]{E^{#1}}
   \nc {\euf}{\eu{4}}
   \nc {\eun}{\eu{n}}
   \nc {\zn}[1]{\intg^{#1}}
   \nc {\zf}{\zn{4}}
   \nc {\zt} {Z_2}
   \nc {\ztn}[1]{\zt^{#1}}
   \nc {\ztt} {\ztn{2}}
   \nc {\ztth} {\ztn{3}}
   \nc {\ztf} {\ztn{4}}
   \nc {\ztnf}{\ztn{n}}
   \nc {\per}[1]{S_{#1}}
   \nc {\pern}{\per{n}}
   \nc {\irrr}[2]{IRR_{#1}({#2})}
   \nc {\irrc}[1]{\irrr{\complex}{#1}}
   \nc {\irrcg}{\irrc{G}}

   \nc {\mn}{(-)}
   \nc {\mns}[1]{\mn^{#1}}
   \nc {\mo}{(-1)}
   \nc {\mos}[1]{\mo^{#1}}
   \nc {\unit} {{\bf 1}}
   \nc {\unt} {\unit_{2\times 2}}
   \nc {\unth} {\unit_{3\times 3}}
   \nc {\unf} {\unit_{4\times 4}}
   \nc {\une} {\unit_{8\times 8}}
   \nc {\unw} {\unit_{12\times 12}}
   \nc {\zrt} {{\bf 0}_{2\times 2}}
   \nc {\zrth} {{\bf 0}_{3\times 3}}
   \nc {\zrf} {{\bf 0}_{4\times 4}}
    \def\CDalign#1{\bgroup\vcenter\bgroup\tabskip 2pt 
      \baselineskip 14pt \lineskip 3pt \lineskiplimit 3pt
      \halign\bgroup &\hfill$##$\hfill\crcr
      #1\crcr\egroup\egroup\egroup} 
    
  \nc{\act}[3]{S_{{#1}}^{{#2}}[{#3}]}
  \nc{\cact}[2]{S_{Cl}^{{#1}}[{#2}]}
  
   \nc {\prj}[4]{\pi_{#1,#2}#2\ot_{S_#1}e^o_{#3,#4}}
   \nc {\prjf}[2]{\prj{o}{#1}{\eta}{#2}}
   \nc {\prjff}{\prjf{h}{i}}
   \nc {\Prj}[4]{\Pi_{#1,#2;#3,#4}}
   \nc {\Prjf}[2]{\Prj{o}{#1}{\eta}{#2}}
   \nc {\Prjff}{\Prjf{h}{i}}
   \nc {\orbt}[2]{\pi_{#1,#2}}
   \nc {\orbtf}[1]{\orbt{o}{#1}}
   \nc {\orbte}{\orbtf{e}}
   \nc {\rp}[3]{\orbt{#1}{#2}(#3)}
   \nc {\rpf}[2]{\rp{o}{#1}{#2}}
   \nc {\stbrp}[2]{D^o_#1(\tld{s}(#2))}
   \nc {\stbrpme}[4]{\stbrp{#1}{#2}^{#3}_{#4}}
   \nc {\stbrpf}[1]{\stbrp{\eta}{#1}}
   \nc {\stbrpmef}[3]{\stbrpme{\eta}{#1}{#2}{#3}}
   \nc {\ch}[2]{\chi_{#1;#2}}
   \nc {\che}[3]{\ch{#1}{#2}(#3)}
   \nc {\chf}[1]{\ch{o}{#1}}
   \nc {\chff}{\chf{\eta}}
   \nc {\chef}[2]{\che{o}{#1}{#2}}
   \nc {\cheff}[1]{\chef{\eta}{#1}}
   \nc {\chsb}[1]{\chi^o_{#1}}
   \nc {\chsbe}[2]{\chsb{#1}(#2)}
   \nc {\chsbf}{\chsb{\eta}}
   \nc {\chsbef}[1]{\chsbe{\eta}{#1}}
   \nc {\cg}[1]{O_{#1}}
   \nc {\cgn} {\cg{n}}
   \nc {\oh} {\cg{4}}
   \nc {\ohd} {{\overline{\oh}}}
   \nc {\ztfb} {\overline{\ztf}}
   \nc {\iso}[2]{ISO_{#1}(#2)}
   \nc {\isod}[1]{\iso{d}{#1}}
   \nc {\eg}[1]{ISO(#1)}
   \nc {\egn}{\eg{n}}
   \nc {\fs}[1]{F(#1)}
   \nc {\fsf}{\fs{\emb}}
   \nc {\fx}[1]{I(#1)}
   \nc {\fxf}{\fx{\emb}}
   \nc {\wrn}[1]{\ztn{#1}\smdp\per{#1}}
   \nc {\wn}{\wrn{n}}
   \nc {\wnn}{\pern^{\zt}}
   \nc {\w}[1]{\per{#1}^{\zt}}
   \nc {\fix}[1]{\per{(n-#1)}\ot \per{p}}
   \nc {\fixp}{\fix{p}}
   \nc {\cb}[1]{C_{#1}}
   \nc {\cn}{\cb{n}}

   \nc {\prm}{\sigma}
   \nc {\prmt}{\tld{\prm}}
   \nc {\prmp}{\prm^\pr}
   \nc {\cyc}[2]{\tau_{{#1}{#2}}}
   \nc {\emb}{\iota}
   \nc {\epy}{\tld{\ep}}
   \nc {\de}[1]{d_{E^{#1}}}
   \nc {\den}{\de{n}}
   \nc {\dy}{\tld{d}}
   \nc {\repw}[2]{e_{(#1)#2}}
   \nc {\repww}[4]{\repw{#1}{#2}\ot\repw{#3}{#4}}
   \nc {\prw}[6]{\pi_{#1,#2}#2\ot_{F_#1}(\repww{#3}{#4}{#5}{#6})}
   \nc {\dm}[1]{d_{({#1})}}
   \nc {\sls}{{\it slash}}
   \nc {\slsv}{{\it slash} }
   \nc {\lcl} {{\it local}}
   \nc {\lclv} {{\it local} }
   \nc {\drln}[5]
    {\put(#1,#2){\line(#3,#4){#5}}}
   \nc {\ybxa}[4]
    {
     \begin{picture}(40,10)
      \drln {0}{0}{0}{1}{10}
      \drln {0}{0}{1}{0}{40}
      \drln {10}{0}{0}{1}{10}
      \drln {0}{10}{1}{0}{40}
      \drln {20}{0}{0}{1}{10}
      \drln {30}{0}{0}{1}{10}
      \drln {40}{0}{0}{1}{10}
      \drln {0}{0}{1}{1}{#1}
      \drln {10}{0}{1}{1}{#2}
      \drln {20}{0}{1}{1}{#3}
      \drln {30}{0}{1}{1}{#4}
     \end{picture}
    }
   \nc {\ybxb}[4]
    {
     \begin{picture}(30,20)
      \put(0,0){\line(0,1){20}}
      \put(0,0){\line(1,0){10}}
      \put(10,0){\line(0,1){20}}
      \put(0,10){\line(1,0){30}}
      \put(0,20){\line(1,0){30}}
      \put(20,10){\line(0,1){10}}
      \put(30,10){\line(0,1){10}}
      \drln {0}{0}{1}{1}{#1}
      \drln {0}{10}{1}{1}{#2}
      \drln {10}{10}{1}{1}{#3}
      \drln {20}{10}{1}{1}{#4}
     \end{picture}
    }
   \nc {\ybxc}[4]
    {
     \begin{picture}(20,20)
      \put(0,0){\line(0,1){20}}
      \put(0,0){\line(1,0){20}}
      \put(10,0){\line(0,1){20}}
      \put(0,10){\line(1,0){20}}
      \put(0,20){\line(1,0){20}}
      \put(20,0){\line(0,1){20}}
      \drln {0}{0}{1}{1}{#1}
      \drln {10}{0}{1}{1}{#2}
      \drln {0}{10}{1}{1}{#3}
      \drln {10}{10}{1}{1}{#4}
     \end{picture}
    }
   \nc {\ybxd}[4]
    {
     \begin{picture}(20,30)
      \drln {0}{0}{0}{1}{30}
      \drln {0}{0}{1}{0}{10}
      \drln {10}{0}{0}{1}{30}
      \drln {0}{10}{1}{0}{10}
      \drln {0}{20}{1}{0}{20}
      \drln {0}{30}{1}{0}{20}
      \drln {20}{20}{0}{1}{10}
      \drln {0}{0}{1}{1}{#1}
      \drln {0}{10}{1}{1}{#2}
      \drln {0}{20}{1}{1}{#3}
      \drln {10}{20}{1}{1}{#4}
     \end{picture}
    }
   \nc {\ybxe}[4]
    {
     \begin{picture}(10,40)
      \drln {0}{0}{0}{1}{40}
      \drln {10}{0}{0}{1}{40}
      \drln {0}{0}{1}{0}{10}
      \drln {0}{10}{1}{0}{10}
      \drln {0}{20}{1}{0}{10}
      \drln {0}{30}{1}{0}{10}
      \drln {0}{40}{1}{0}{10}
      \drln {0}{0}{1}{1}{#1}
      \drln {0}{10}{1}{1}{#2}
      \drln {0}{20}{1}{1}{#3}
      \drln {0}{30}{1}{1}{#4}
     \end{picture}
    }
   
   \nc{\op}{orientation-preserved}
   \nc{\opv}{orientation-preserved }
   \nc{\on}[1]{SO_{#1}}
   \nc{\ohn}[1]{O_{#1}}
   \nc{\of}{\on{4}}
   \nc{\ohf}{\ohn{4}}
   \nc{\ofd}{\overline{\of}}
   \nc{\ohfd}{\overline{\ohf}}
   \nc{\cir}[1]{e^{i({#1})\pi}}
   
   \def\pthv{Pythagoras' Theorem }
   \def\dop{Dirac operator}
   \def\dopv{Dirac operator }

   \nc {\norm}[1]{\parallel{#1}\parallel}

   \nc{\groupsum}[1]{{\sum_{#1}}^{\pr}}
   \nc{\reducedsum}[1]{{\sum_{#1}}^{\pr\pr}}
 \title{\pthv on a 2D-Lattice from a ``Natural'' Dirac Operator and Connes'
 Distance Formula
 }
 \author{Jian Dai\thanks{daijianium@yeah.net},
  Xing-Chang Song\thanks{songxc@ibm320h.phy.pku.edu.cn}\\
  Theoretical Group, Department of Physics\\
  Peking University
  Beijing, P. R. China, 100871}
 \date{December 14th, 2000}
 
\begin{document}
  \maketitle
  \begin{abstract}
  \noindent
   One of the key ingredients of A. Connes' noncommutative geometry is a generalized
   \dopv which induces a metric(Connes' distance) on the state space. We generalize such a \dopv devised by A. Dimakis \etal,
   whose Connes' distance recovers the linear distance on a 1D lattice, into 2D lattice. This \dopv being
   ``naturally'' defined has the ``local eigenvalue property'' and
   induces Euclidean distance on this 2D lattice. This kind of \dopv can
   be generalized into any higher dimensional lattices.\\

  \end{abstract}
  \section{Introduction}
   Lattice \dopv is a long-standing problem embarrassing
   lattice field theorists. No-Go theorem \cite{nielsen} makes the implementation of
   chiral fermions on lattices almost impossible. Recent years, some
   breakthroughs have been achieved, e.g. the rediscovery of Ginsparg-Wilson relation
   \cite{luscher}, the devices of domainwall and overlap \dop s \cite{domainwall}\cite{overlap}.
   On the other hand, lattices can be considered as a simplest realization of
   noncommutative geometry (NCG) which has drawn
   more and more attention of theoretical physicists due to
   its applications in standard model of particle physics \cite{martin}\cite{carminati}\cite{iochum0},
   lattice field theory \cite{dimakis3}\cite{dimakis4}\cite{dimakis1},
   and string/M-theory \cite{douglas}\cite{seiberg}. NCG provides a powerful candidate
   of mathematical framework for geometrical understanding of fundamental physical laws.
   In Alain Connes's version of NCG, a generalized \dopv plays a central
   role in introducing the metric structure onto a noncommutative space
   \cite{connes}\cite{connes0}\cite{connes1}. Then intuitively,  it becomes an interesting question whether
   Connes' NCG idea could brighten the problem of lattice \dop.
   In fact, some groups have explored this question. G.
   Bimonte \etal first pointed out that na\"{\i}ve \dopv is not able to
   induce the conventional distance on a 4D lattice by Connes' construction and Wilson-\dopv
   gives an even worse result \cite{bimonte}. Starting from this observation, E. Atzmon computed
   this ``anomalous distance'' induced by na\"\i ve \dopv for an 1D lattice precisely
   \cite{atzmon}, which gives
   \[
    d(0,2n-1)=2n, d(0,2n)=2\sqrt{n(n+1)} (n\in \natural)
   \]
   Hence, we can say that NCG
   provides another criterion for the rational choices of \dopv on lattices.
   A. Dimakis \etal discovered an 1D operator whose Connes' distance
   coincides with the usual linear distance, when the number of lattice sites is finite \cite{dimakis2}.
   This serial work indicates that it is a highly nontrivial problem to devise a proper \dopv
   whose Connes' distance is a desired one.\\

   In this paper, we construct a ``natural'' \dopv on a 2D
   lattice. The induced metric of this operator endows conventional Euclidean geometry on this lattice,
   i.e. the \pthv holds for the Connes' distance. The restoration
   of metric relies on that this operator has a so-called ``local eigenvalue property'', such that
   the norm $\norm{[D,f]}$ is completely solvable (diagonalizable, integrable).
   The \dopv devised by us can be regarded as a generalization of Dimakis' operator in 1D case;
   it can be generalized into
   any high dimensional lattices easily.\\

   The paper is organized as follows. In section \ref{sec1}, we give a brief introduction to Connes'
   NCG and his distance formula.
   The ``natural'' \dopv is defined in section \ref{sec2}.
   In section \ref{sec3}, we explore the ``local eigenvalue
   property'' of our \dop.
   In section \ref{sec4}, we prove that our \dopv implies the \pthv on a
   2D Lattice.
   In section \ref{sec5}, we generalize the ``natural'' \dopv into
   any higher dimensional lattices and make some open discussions.
  \section{Noncommutative Geometry and Connes' Distance
  Formula}\label{sec1}
   The object of a Connes' NCG is a triple $(A,H,D)$ called a K-cycle, in which $A$
   is an involutive algebra, represented faithfully and unitarily as a subalgebra of
   bounded operators on a Hilbert space $H$, and $D$ is a self-adjoint operator on $H$,
   which is called generalized \dop, with compact resolvent so that $[D,\hat{a}]$
   is bounded for all $a$ in $A$. Here we use $\hat{a}$ to denote the
   imagine of $a$ on $H$; without introducing any misunderstanding
   below, we will just omit the hat on $a$.
   A K-cycle is required to satisfy some
   axioms such that it recovers the ordinary spin geometry
   on a differential manifold when $A$ is taken to be the algebra of smooth functions
   over this spin manifold, $H$ is the space of $L^2$-spinors and
   $D$ is the classical \dop \cite{connes1}. \\

   From a K-cycle, we can define a metric $d_D(,)$ on the state space of
   $A$ denoted as $S(A)$, which we have referred as induced metric or Connes'
   distance.
   \eq
   \label{dis1}
    d_D(\phi, \psi)=sup\{|\phi (a)-\psi (a)|/ a\in A, \norm{[D,a]}\leq 1\}
   \en
   for any $\phi, \psi \in S(A)$. One can check that this definition satisfies all
   of the three axioms for a metric easily. Once $A$ is commutative, pure
   states correspond to characters which can be interpreted as
   points, with $A$ being the algebra of functions over these
   points (Gel'fand-Naimark Theorem); and Eq.(\ref{dis1}) can, under
   this circumstance, be
   rewritten as
   \eq
   \label{dis2}
    d_D(p, q)=sup\{|f(p)-f(q)|/f\in A, \norm{[D,f]}\leq
    1\}, \forall p,q\in S(A)
   \en
   Noticing the fact that
   \[
    \norm{a}^2=sup\{(a\psi,a\psi)/\psi\in H,
    \norm{\psi}=1\}=sup\{\lambda(a)\mid a^\ast
    a\psi=\lambda(a)\psi\}, a\in B(H)
   \]
   where $B(H)$ is the algebra of bounded operators on $H$,
   we can express the inequality constraint in Eq.(\ref{dis2}) by using eigenvalues
   \eq
   \label{dis3}
    d_D(p, q)=sup\{|f(p)-f(q)|/f\in A, \lambda([D,f]^\dag [D,f])\leq
    1\}, \forall p,q\in S(A)
   \en
   {\bf Notations}\\
   $\sigma_i, i=1,2,3$ are Pauli matrices defined in the
   ordinary way. If $S$ is a finite set, then $|S|$ is the number of the
   elements in $S$. Let $i=1,2,3$, define $\gamma$-matrices as
   \[\gamma^i=
    \left(
     \begin{array}{cc}
      &\sigma_i\\
      \sigma_i&
     \end{array}
    \right);
    \gamma^4=
    \left(
     \begin{array}{cc}
      &i\\
      -i&
     \end{array}
    \right)
   \]
   The vacant matrix elements are understood as zeros and this
   will be taken as a convention all through this paper. Introduce
   $\gamma^1_\pm={1\over 2}(\gamma^1 \pm i\gamma^2),
   \gamma^2_\pm={1\over 2}(\gamma^3 \pm i\gamma^4)$, whose explicit
   matrix representations are
   \[
    \gamma^1_+=
     \left(
      \begin{array}{cccc}
       &&&1\\&&0&\\&1&&\\0&&&
      \end{array}
     \right);
    \gamma^1_-=
     \left(
      \begin{array}{cccc}
       &&&0\\&&1&\\&0&&\\1&&&
      \end{array}
     \right);
   \]
   \[
    \gamma^2_+=
     \left(
      \begin{array}{cccc}
       &&0&\\&&&-1\\1&&&\\&0&&
      \end{array}
     \right);
    \gamma^2_-=
     \left(
      \begin{array}{cccc}
       &&1&\\&&&0\\0&&&\\&-1&&
      \end{array}
     \right)
   \]
   satisfying Clifford algebra relations
   \[
    \{\gamma^i_\pm,\gamma^j_\pm\}=0,
    \{\gamma^i_\pm,\gamma^j_\mp\}=\dlt^{ij}, i,j=1,2
   \]
   $\intg$ is referred to the set of integrals and $\complex$ is
   for complex numbers. We adopt the convention $a=1$ for the
   lattice constant $a$ in this paper.
  \section{``Natural'' Dirac Operator on a 2D Lattice}\label{sec2}
   First we give a detailed re-formulation of Dimakis' operator on the 1D lattice.
   Let $L_1=\{x|x\in\intg\}$, $A_0=\{f:L_1\rightarrow \complex\},
   H=l^2(L_1)\otimes\complex^2=\{\psi\}$.
   The operator developed by Dimakis \etal acting on $H$ can be written in the form
   \[
    D=\sigma_{+}\partial^{+}+\sigma_{-}\partial^{-}
    =\left(
      \begin{array}{cc}
       &\partial^+\\\partial^-&
      \end{array}
     \right)
   \]
   where $\sigma_\pm=(\sigma_1 \pm i\sigma_2)/2$, $(T^\pm\psi)(x):=\psi(x\pm 1)$,
   $(\partial^\pm\psi)(x)=((T^\pm -\unit)\psi)(x)=\psi(x\pm
   1)-\psi(x)$ and $\psi(x)=(\psi_1,\psi_2)^T(x)$. Refine $A_0$ to be $A=\{f\in A_0/
   \norm{[D,f]}<\infty\}$.\\

   We point out that $D$ has a ``local eigenvalue property'', showing below
   \[
    [D,f]=
    \left(
     \begin{array}{cc}
      &(\partial^+f)T^+\\
      (\partial^-f)T^-&
     \end{array}
    \right)
    =-
    \left(
     \begin{array}{cc}
      &T^+\cdot (\partial^-f)\\
      T^-\cdot(\partial^+f)&
     \end{array}
    \right)
   \]
   \[
    [D,f]^\dag=-[D,\bar{f}]
   \]
   Introduce a ``Hamiltonian''$H(df):=[D,f]^\dag [D,f]$
   \[=
     \left(
      \begin{array}{cc}
       |\partial^+f|^2&\\&|\partial^-f|^2
      \end{array}
     \right)
   \]
   Consider the eigenvalue equation
   \[
    H(df)\psi=\lambda(df)\psi\Leftrightarrow
   \]
   \[
    |\partial^+f|^2(x)\psi_1(x)=\lambda\psi_1(x), \forall x\in
    \intg
   \]
   \[
   |\partial^-f|^2(x)\psi_2(x)=\lambda\psi_2(x), \forall x\in
   \intg
   \]
   So that the constraint in Eq.(\ref{dis3}) implies
   \eq
   \label{1d1}
    \lambda(x,1)=|\partial^+f|^2(x)\leq 1, \forall x\in \intg
   \en
   \eq\label{1d2}
    \lambda(x,2)=|\partial^-f|^2(x)\leq 1, \forall x\in \intg
   \en
   Now we observe that each eigenvalue of $H(df)$ is just related to one
   link
   of $L_1$, to which we refer as ``local eigenvalue''. As a comparison, the
   eigenvalue equation for 1D na\"{\i}ve \dopv $D_N:=(T^+-T^-)/2$
   is
   \[
    {1\over 4}(|\partial^+f|^2(x)+ |\partial^-f|^2(x))\psi(x) +
    {1\over 4}((\partial^+\bar{f})(x)(\partial^+f)(x+1)\psi(x+2)+
    (\partial^-\bar{f})(x)(\partial^-f)(x-1)\psi(x-2))=\lambda\psi(x)
   \]
   for all x in $\intg$,
   with $A_N=A$, $H_N=l^2(L_1)$, which possesses no ``local eigenvalues" evidently.\\

   Now we induce $d_D(,)$ on $L_1$. Let $f\in A$ subjected to (\ref{1d1})(\ref{1d2}), i.e.
   $\norm{[D,f]}\leq 1$, so that $|f((x+m)-f(x)|\leq m,
   m=1,2,3,...$; hence, $d_D(x+m,x)$ has an upper bound $m$. Let $f_0(x)=x$, then
   $f_0$ saturates this upper bound. Thus, we have proved $d_D(m,n)=|m-n|,
   \forall m,n\in L_1$.\\

   The 2D lattice being considered as a set is parametrized as $L_2=\{x=(m,n)\mid m,n\in \intg\}$.
   \[
    A_0:=\{f:L_2\rightarrow \complex\}, H:=l^2(L_2)\otimes
    \complex^4=\{\psi :\psi=(\psi_1,\psi_2,\psi_3,\psi_4)^T\}
   \]
   \eq\label{dirac}
    D=\sum_{i=1}^{2}{\sum_{s=\pm}{\gamma^i_s\partial^s_i}}
     =\left(
      \begin{array}{cccc}
       &&\partial^-_2&\partial^+_1\\
       &&\partial^-_1&-\partial^+_2\\
       \partial^+_2&\partial^+_1&&\\
       \partial^-_1&-\partial^-_2&&
      \end{array}
      \right)
   \en
   \[
    (T^\pm_i\psi)(x)=\psi(x\pm \hat{i})
   \]
   \[
    (\partial^\pm_i\psi)(x)=((T^\pm_i -\unit)\psi)(x)=\psi(x\pm \hat{i})-\psi(x)
   \]
   where $\hat{1}=(1,0), \hat{2}=(0,1)$. Refine $A_0$ to be $A:=\{f\in A_0/
   \norm{[D,f]}<\infty\}$. By symmetries of $D$, we have
   \[
    d_D((m,n),(m^\pr,n^\pr))=d_D((0,0),(m^\pr-m,n^\pr-n))
   \]
   so that to consider $d_D((0,0),(m,n))$ is enough. What's
   more,
   \[
    d_D((0,0),(m,n))=d_D((0,0),(-n,m)), d_D((0,0),(m,n))=d_D((0,0),(n,m)),
   \]
   so we just need to consider $m\geq n\geq 0$.\\

   To end this section , we adopt the method in \cite{bimonte} to prove
   \eq\label{one}
    d_D((0,0),(i,0))=i, i=1,2,...
   \en
   \begin{lemma}
    Let $f\in A$, $\norm{[D,f]}\leq 1$ and $\tld{f}(m,n):=f(m,0)$, then
    $\norm{[D,\tld{f}]}\leq 1$.
   \end{lemma}
   \prf
    By the definition of $D$ in (\ref{dirac}), there is
    \[
     [D,\tld{f}]\psi=((\partial^+_1\tld{f})(T^+_1\psi_4),
     (\partial^-_1\tld{f})(T^-_1\psi_3),(\partial^+_1\tld{f})(T^+_1\psi_2),(\partial^-_1\tld{f})(T^-_1\psi_1))^T
    \]
    Thus, notice the definition of $\tld{f}$,
    \[
     ([D,\tld{f}]\psi,[D,\tld{f}]\psi)=\sum_{m,n}{(|\partial^+_1\tld{f}|^2(m,n)(\sum_{i=1}^4
     {|\psi^\pr_i|^2(m,n)}))}
    \]
    \[
     =\sum_m{(|\partial^+_1f|^2(m,0)(\sum_n\sum_{i=1}^4
     {|\psi^\pr_i|^2(m,n)}))}
    \]
    \eq\label{2to1}
     \Rightarrow
     \norm{[D,\tld{f}]}=sup\{|\partial^+_1f|(m,0)/m\in\intg \}<\infty
    \en
    Define $\hat{H}:=\{\psi\in H/ \psi_i=0, i=1,2,3\}$, then
    \eq
    \label{ineq}
     \norm{[D,f]}_{\hat{H}}\leq \norm{[D,f]}\leq 1
    \en
    and
    \[
     ([D,f]\psi,[D,f]\psi)|_{\hat{H}}=\sum_{m,n}{|\partial^+_1f|^2(m,n)|\psi_4|^2(m+1,n)
     +|\partial^+_2f|^2(m,n)|\psi_4|^2(m,n+1)}
    \]
    For any $\psi\in H$, we define $\hat{\psi}\in \hat{H}$ by that
    $\hat{\psi}_4(m,n)=0, n\neq 0$ and that $\hat{\psi}_4(m,0)$ satisfy\\
    $|\hat{\psi}_4(m,0)|^2=\sum_n\sum_{i=1}^4
    {|\psi^\pr_i|^2(m,n)}$, for all $m$. Therefore,
    \[
     ([D,\tld{f}]\psi,[D,\tld{f}]\psi)
     =\sum_m{(|\partial^+_1f|^2(m,0)|\hat{\psi}_4(m,0)|^2}
     \leq([D,f]\hat{\psi},[D,f]\hat{\psi})|_{\hat{H}}\leq \norm{[D,f]}_{\hat{H}}
    \]
    Notice (\ref{ineq}), and there are
    \[
     \norm{[D,\tld{f}]}\leq \norm{[D,f]}_{\hat{H}}\leq \norm{[D,f]}\leq 1
    \]
   \endprf
   Following this lemma, we can just consider $f$ with
   $\partial^+_2f=0$ and reach the conclusion $d_D((0,0),(i,0))\leq i,
   i=1,2,...$; $f_0(m,n)=m$ saturates this upper bound.
   Hence, Eq.(\ref{one}) holds.
  \section{Local Eigenvalues}\label{sec3}
   We have to consider the eigenvalue problem in Eq.(\ref{dis3}),
   when discussing distance $d_D((0,0),(m,n)), m\geq n\geq 1$.
   Fortunately, our ``natural'' \dopv has the ``local eigenvalue
   property'' also. Here we give a very detailed calculation.\\

   Let $D(f):=[D,f]$
   \[=
    \left(
     \begin{array}{cccc}
      &&(\partial^-_2f)T^-_2&(\partial^+_1f)T^+_1\\
      &&(\partial^-_1f)T^-_1&-(\partial^+_2f)T^+_2\\
      (\partial^+_2f)T^+_2&(\partial^+_1f)T^+_1&&\\
      (\partial^-_1f)T^-_1&-(\partial^-_2f)T^-_2&&
     \end{array}
    \right)
   \]
   One can check $[D,f]^\dag=-D(\bar{f})$. Define the
   ``Hamiltonian" $H(df):=[D,f]^\dag [D,f]=-D(\bar{f})D(f)$
   \[=
    \left(
     \begin{array}{cc}
      |\partial^+_1f|^2+|\partial^-_2f|^2&
      (T^-_2(\partial^+_2\bar{f} \partial^+_1f)-
      T^+_1(\partial^-_1\bar{f}\partial^-_2f))T^+_1T^-_2\\
      (T^-_1(\partial^+_1\bar{f} \partial^+_2f)-
      T^+_2(\partial^-_2\bar{f}\partial^-_1f))T^-_1T^+_2&|\partial^-_1f|^2+|\partial^+_2f|^2
     \end{array}
    \right)\oplus
   \]
   \[
    \left(
     \begin{array}{cc}
      |\partial^+_1f|^2+|\partial^+_2f|^2&
      (T^+_2(\partial^-_2\bar{f} \partial^+_1f)-
      T^+_1(\partial^-_1\bar{f}\partial^+_2f))T^+_1T^+_2\\
      (T^-_1(\partial^+_1\bar{f} \partial^-_2f)-
      T^-_2(\partial^+_2\bar{f}\partial^-_1f))T^-_1T^-_2&|\partial^-_1f|^2+|\partial^-_2f|^2
     \end{array}
    \right)
   \]
   with the eigenvalue equation
   \eq\label{pqeqn}
    H(df)\psi=\lambda(df)\psi
   \en
   Fortunately, Eq.(\ref{pqeqn}) can be reduced to a collection of
   equation sets, with each equation set being related to a fundamental plaque $\{(m,n),(m+1,n),
   (m,n-1),(m+1,n-1)\}\subset L_2$
   \eq\label{pq1}
    \left(
     \begin{array}{cc}
      \rho_1^2 +\rho_4^2-\lambda&
      \overline{\Delta_4}\Delta_3-\overline{\Delta_1}\Delta_2\\
      \Delta_4\overline{\Delta_3}-\Delta_1\overline{\Delta_2}&\rho_2^2 +\rho_3^2-\lambda
     \end{array}
    \right)
    \left(
     \begin{array}{c}
      \psi_1(m,n)\\\psi_2(m+1,n-1)
     \end{array}
    \right)=0
   \en
   \eq\label{pq2}
    \left(
     \begin{array}{cc}
      \rho_3^2 +\rho_4^2-\lambda&
      -\overline{\Delta_4}\Delta_1+\overline{\Delta_3}\Delta_2\\
      -\Delta_4\overline{\Delta_1}+\Delta_3\overline{\Delta_2}&\rho_1^2 +\rho_2^2-\lambda
     \end{array}
    \right)
    \left(
     \begin{array}{c}
      \psi_3(m,n-1)\\\psi_4(m+1,n)
     \end{array}
    \right)=0
   \en
   in which $\Delta_1(m,n):=(\partial^+_1f)(m,n)$,
   $\Delta_2(m,n):=(\partial^+_2f)(m+1,n-1)$,
   $\Delta_3(m,n):=(\partial^+_1f)(m,n-1)$,
   $\Delta_4(m,n):=(\partial^+_2f)(m,n-1)$ and, without misleading, we have omitted the arguments
   $(m,n)$. Let $\Delta_i:=\rho_i e^{i\theta_i}, i=1,2,3,4$,
   $\theta:=\theta_1+\theta_3-\theta_2-\theta_4$, and
   \[
    A:=\rho_1^2 +\rho_4^2, B:=\rho_2^2 +\rho_3^2, C:=\overline{\Delta_4}\Delta_3-\overline{\Delta_1}\Delta_2
   \]
   \[
    A^\pr:=\rho_3^2 +\rho_4^2, B^\pr:=\rho_1^2 +\rho_2^2,
    C^\pr:=-\overline{\Delta_4}\Delta_1+\overline{\Delta_3}\Delta_2
   \]
   then the secular equation for Eq.(\ref{pq1}) is
   \[
    (\lambda-A)(\lambda-B)-C\bar{C}=\lambda^2-(A+B)\lambda+AB-C\bar{C}=0
   \]
   in which
   \[
    C\bar{C}=\rho_3^2\rho_4^2+\rho_1^2\rho_2^2-2\rho_1\rho_2\rho_3\rho_4\cos{\theta}
   \]
   The solutions are
   \eq\label{l1}
    \lambda_\pm={1\over 2}({A+B\pm\sqrt{(A-B)^2+4C\bar{C}}})
   \en
   Similarly for Eq.(\ref{pq2}),
   \eq\label{l2}
    \lambda^\pr_\pm={1\over 2}(A^\pr+B^\pr\pm\sqrt{(A^\pr-B^\pr)^2+4C^\pr\bar{C^\pr}})
   \en
   in which
   \[
    C^\pr\bar{C^\pr}=\rho_1^2\rho_4^2+\rho_2^2\rho_3^2-2\rho_1\rho_2\rho_3\rho_4\cos{\theta}
   \]
   Therefore, $\norm{[D,f]}\leq 1\Leftrightarrow \forall (m,n)$,
   \eq\label{eigen}
    \lambda_+(m,n)\leq 1,\lambda^\pr_+(m,n)\leq 1
   \en
   Equivalently, Eqs.(\ref{l1})(\ref{l2}) and inequality (\ref{eigen})
   $\Leftrightarrow$
   \eq\label{link1}
    1+\rho_1^2\rho_3^2+\rho_2^2\rho_4^2+2\rho_1\rho_2\rho_3\rho_4\cos{\theta}\geq
    \rho_1^2+\rho_2^2+\rho_3^2+\rho_4^2
   \en
   \eq\label{link2}
    2\geq\rho_1^2+\rho_2^2+\rho_3^2+\rho_4^2
   \en
   There is another constraint, the closedness condition
   \eq\label{link3}
    \Delta_1+\Delta_4=\Delta_2+\Delta_3
   \en
   Hence we get (\ref{link1})(\ref{link2})(\ref{link3}) as the
   specific expressions of $\norm{[D,f]}\leq 1$ on $L_2$.\\

   To end this section, we give Eq.(\ref{one}) a second proof using
   (\ref{eigen}). In fact, just notice
   \[
    \rho_1^2\leq\rho_1^2+\rho_4^2\leq \lambda_+\leq 1
   \]
   and we have $\rho_i\leq 1, i=1,2,3,4$. Consequently, $d_D((0,0),(i,0))$
   has the upper bound $i$.
  \section{\pthv for $d_D(,)$ on $L_2$}\label{sec4}
   Let $m\geq n\geq 1$ below and we claim that
   \begin{theorem}\label{thph}(\pthv on 2D lattice)
    \eq\label{pth}
     d_D((0,0),(p,q))=\sqrt{p^2+q^2}, \forall p\geq q\geq 1
    \en
   \end{theorem}
   The proof needs only inequalities (\ref{link2})(\ref{link3}) and we will treat $q=1$ and $q=2,3,4,...$
   respectively. We need an important inequality in mathematics
   analysis
   \eq\label{ine}
    \sum_{i=1}^n{a_i}\sum_{i=1}^n{b_i}\leq n\sum_{i=1}^n{a_ib_i}
   \en
   where the equality holds if $a_1=a_2=...=a_n$ or
   $b_1=b_2=...=b_n$.\\

   The general philosophy of the proof can be illustrated as
   following. First we define the concept of {\it canonical path},
   which means a subset of $L_2$ generated by one point denoted as {\it start}
   and the operations $T^+_1, T^+_2$. Then it must have a {\it end} if
   $T^+_1, T^+_2$ just act finite times. Second we pick out a rectangle subset $L(p,q)$ of $L_2$ which is defined
   as $\{(m,n)/m=0,1,...,p, n=0,1,..,q\}$, and define the link set
   on $L(p,q)$, $B(p,q):=\{(m,n,s)/s=1,2;s=1:m=0,1,...,p-1,n=0,1,...,q;
   s=2:
   m=0,1,...,p,n=0,1,...,q-1\}$. For any $f|_{L(p,q)}$ defined on $L(p,q)$, there is a
   function $\Delta f$ defined on $B(p,q)$ correspondingly, with
   $\Delta f(m,n,1)=(\partial^+_1 f)(m,n), \Delta f (m,n,2)=(\partial^+_2
   f)(m,n)$. There is another set induced by $L(p,q)$, the set of
   all canonical paths starting at $(0,0)$ and ending at $(p,q)$, written as
   $\Gamma(p,q)$. Any element $\gamma\in \Gamma(p,q)$ can be considered
   as $p+q$ sequential links, namely a subset of $B(p,q)$. As for
   the first step of proof, we want to show for any $\gamma\in
   \Gamma(p,q)$, $|\sum_{l\in \gamma}{\Delta f
   (l)}|\leq\sqrt{p^2+q^2}$, if $f$ subjects to (\ref{link2})(\ref{link3}).
   Adopting reduction of absurdity, we suppose there is
   a $\gamma_0$ such that $|\sum_{l\in \gamma_0}{\Delta f
   (l)}|>\sqrt{p^2+q^2}$. Then by virtue of the closedness condition (\ref{link3}),
   $|\sum_{l\in \gamma}{\Delta f
   (l)}|>\sqrt{p^2+q^2}, \forall \gamma\in \Gamma(p,q)$, which implies
   \[
    \sum_{l\in \gamma}{|\Delta f(l)|^2}>\frac{p^2+q^2}{p+q}
   \]
   by (\ref{ine}).
   Therefore
   \eq\label{bd}
    \sum_{\gamma\in\Gamma(p,q)}(\sum_{l\in \gamma}{\Delta
    f(l)^2})> R(p,q)\frac{p^2+q^2}{p+q}
   \en
   where we introduce
   $R(p,q)=\sum_{\gamma\in\Gamma(p,q)}1=|\Gamma(p,q)|$, the
   number of all canonical paths starting from $(0,0)$ and ending
   $(p,q)$, which is equal to the binomial coefficient
   $\left(\begin{array}{c}p+q\\p\end{array}
    \right)$. We change the summation for paths in $\Gamma(p,q)$ in (\ref{bd}) to
    the summation in $B(p,q)$,
    \eq\label{note1}
     \sum_{l\in B(p,q)}(L(l)\Delta
    f(l)^2)> R(p,q)\frac{p^2+q^2}{p+q}
    \en
    where $L(l)$ is the weight function on $B(p,q)$ introduced by
    the changing of indice of summations. The geometric
    interpretation of $L(l)$ is the number of canonical paths in
    $\Gamma(p,q)$ passing through link $l$, which is expressed as
    \[
     L(m,n,1)=R(m,n)R(p-m-1,q-n), L(m,n,2)=R(m,n)R(p-m,q-n-1)
    \]
    We induce the third set from $L(p,q)$, the set $Q(p,q)$ of all
    fundamental plaques as subsets in $L(p,q)$, which can be
    expressed as $\{(m,n)/m=0,1,...,p-1, n=0,1,...,q-1\}$.
    Rewrite (\ref{link2}) as
    \eq\label{qq}
     \sum_{i=1}^4|\Delta f|^2(p,i)\leq 2
    \en
    which can be considered as a constraint on the four
    links in one fundamental plaque $p$. Now sum (\ref{qq}) for all plaques in
    $Q(p,q)$
    \[
     \sum_{p\in Q(p,q)}\sum_{i=1}^4|\Delta f|^2(p,i)\leq
     2|Q(p,q)|=2pq
    \]
    Again we change the summation for plaques to the summation for
    links and introduce another weight function $S(l)$,
    \eq\label{qq1}
     \sum_{l\in B(p,q)}S(l)|\Delta f|^2(l)\leq 2|Q(p,q)|
    \en
    where $S(m,0,1)=S(m,q,1)=S(0,n,2)=S(p,n,2)=1, S(other)=2$,
    namely to links $l$ shared by two plaques $S(l)=2$, else
    $S(l)=1$.
    Using (\ref{ine}) again and noticing $L(l)\neq 0$, we get
    \[
     2|Q(p,q)|\geq \sum_{l\in B(p,q)}S(l)|\Delta f|^2(l)\geq
     \sum_{l\in B(p,q)}(\frac{S(l)}{L(l)})(L(l)|\Delta f|^2(l))
    \]
    \[
     \geq\frac{1}{|B(p,q)|}\sum_{l\in B(p,q)}(\frac{S(l)}{L(l)})\sum_{l\in B(p,q)}(L(l)|\Delta f|^2(l))
    \]
    \eq\label{note2}
     \Leftrightarrow \frac{2|Q(p,q)||B(p,q)|}{\sum_{l\in
     B(p,q)}\frac{S(l)}{L(l)}}\geq\sum_{l\in B(p,q)}(L(l)|\Delta f|^2(l))
    \en
    where $|B(p,q)|=2pq+p+q$.
    The contradiction lays between inequalities (\ref{note1}) and
    (\ref{note2}), if only
    \begin{lemma}\label{lemma1}
     \eq\label{note3}
      \frac{2|Q(p,q)||B(p,q)|}{\sum_{l\in
     B(p,q)}\frac{S(l)}{L(l)}}\leq R(p,q)\frac{p^2+q^2}{p+q}
     \en
    \end{lemma}
    As we declared that the description above is just general philosophy of proof,
    since the operating on $L(p,q)$ is complicated, we will just prove
    $q=1$ case in Lemma \ref{lemma1} and leave $q\geq2$ cases to
    a revised $L(p,q)$ construction.\\

    As for $q=1$, (\ref{note3})$\Leftrightarrow$
    \eq\label{note4}
     2p(1+3p)\leq (1+p^2)\sum_{l\in
     B(p,1)}\frac{S(l)}{L(l)}
    \en
    Using the definitions of $L(l)$ and $S(l)$, we have
    $\sum_{l\in
     B(p,1)}\frac{S(l)}{L(l)}=2(p+\sum_{n=1}^p {1\over n})$.
    Then (\ref{note4}) reduces to show
    \eq\label{note5}
     3p^2\leq p^3 + (1+p)(\sum_{n=1}^p {1\over n})
    \en
    However, (\ref{note5}) is easily checked by $p=1$, $p=2$, and
    $p=3,4,...$ respectively. Accordingly, we reach the conclusion
    $d_D((0,0),(p,1)), p=1,2,3,...$ has an upper bound
    $\sqrt{1+p^2}$.\\

    We define $\tld{L}(p,q)$ to be a ``folding ruler'', $\{(m,0),(m,1),(p-1,n),(p,n)/m=0,1,..,p,n=2,...,q
    \}$, for $p\geq q\geq 2$, and apply our general philosophy of proof to $\tld{L}(p,q)$.
    Namely, we induce $\tld{B}(p,q), \tld{P}(p,q), \tld{Q}(p,q)$
    from $\tld{L}(p,q)$, and introduce weight functions $\tld{L}(l), \tld{S}(l)$ on
    $\tld{B}(p,q)$. If
    \begin{lemma}\label{lemma2}
     \eq\label{fold1}
      \frac{2|\tld{Q}(p,q)||\tld{B}(p,q)|}{\sum_{l\in
     \tld{B}(p,q)}\frac{\tld{S}(l)}{\tld{L}(l)}}\leq \tld{R}(p,q)\frac{p^2+q^2}{p+q}
     \en
    \end{lemma}
    holds, then we find a contradiction and conclude that $d_D((0,0),(p,q))$ has an upper
    bound $\sqrt{p^2+q^2}$ for all $p\geq q\geq 2$.
    And here, $|\tld{Q}(p,q)|=p+q-1$, $|\tld{B}(p,q)|=3p+3q-2$,
    $\tld{R}(p,q)=1+pq$; $\tld{S}(m,0,2)=\tld{S}(p-1,n,1)=2, m=1,2,...,p-1, n=1,...,q-1, \tld
    {S}(other)=1$; $\tld{L}(m,0,1)=1+q(p-m-1), \tld{L}(m,0,2)=q, m=0,1,...p-1$;
    $\tld{L}(p,0,2)=1$; $\tld{L}(m,1,1)=q(m+1), m=0,1,...p-2$;
    $\tld{L}(p-1,n,2)=p(q-n),\tld{L}(p,n,2)=1+pn, n=1,...,q-1$;
    $\tld{L}(p-1,n,1)=p,n=1,...,q$.\\

    Thus,
    \[
     \sum_{l\in\tld{B}(p,q)}\frac{\tld{S}(l)}{\tld{L}(l)}=2+2({p\over
     q} + {q\over p}) + \dlt\geq 6
    \]
    in which
   \[
    \dlt={1\over q}(\sum_{n=1}^{p-1}{1\over n} -1)+
    {1\over p}(\sum_{n=1}^{q-1}{1\over n} -1)+
    \sum_{n=1}^{p-1}{1\over {1+nq}}+
    \sum_{n=1}^{q-1}{1\over {1+np}}\geq 0
   \]
   We claim Lemma \ref{lemma2} is implied by
   \[
    (p+q)|\tld{Q}(p,q)||\tld{B}(p,q)|\leq 3\tld{R}(p,q)(p^2+q^2)
   \]
   \eq\label{fin}
    \Leftrightarrow (p+q)(p+q-1)(3(p+q)-2)\leq 3(1+pq)(p^2+q^2),
    \forall p\geq q\geq 2
   \en
   We prove (\ref{fin}) by induction. First, fixing one $q=2,3,...$, we show that (\ref{fin}) holds
   when $p=q$. In this case, (\ref{fin}) is reduced to
   \[
    12q^3+2q\leq 3q^4+13q^2, \forall q\geq 2
   \]
   The above inequality can be checked to be valid when $q=2$, $q=2$, and $q\geq
   3$. Second, with this fixed $q$, we suppose that the statement is valid in the case
   $p=p_0$, and show that it will also hold in $p=p_0+1$. Without misunderstanding, we drop
   the subscript ``$0$''. It is sufficient to check
   \[
    9pq+9q^2\leq 9(q-1)p^2+3q^3+7p+4q+3
   \]
   Since $9pq\leq 9(q-1)p^2, \forall p\geq q\geq 2$, we just check
   \[
    9q^2\leq 3q^3+7p+4q+3
   \]
   which is valid when $q=2$ and $q\geq 3$. So the upper bound
   $\sqrt{p^2+q^2}$ follows.\\

   We choose $f_{(p,q)}(m,n)=\frac{pm+qn}{\sqrt{p^2+q^2}}$ which can be checked easily
   to saturate this upper bound. Therefore, theorem \ref{thph}
   follows.
  \section{Discussions}\label{sec5}
   Our ``natural'' \dopv is able to be generalized easily into any
   dimension greater than two. Let $\Gamma^i, i=1,2,...,2d$ be the
   generators of $Cl(E^{2d})$ which satisfy
   \eq\label{clff}
    \{\Gamma^i, \Gamma^j\}=2\dlt^{ij}, i,j=1,2,...,2d
   \en
   Define $\Gamma^k_\pm=(\Gamma^{2k-1}\pm i\Gamma^{2k})/2,
   k=1,2,...,d$, and (\ref{clff}) changes the form into
   \[
    \{\Gamma^m_\pm,\Gamma^n_\pm\}=0,
    \{\Gamma^m_\pm,\Gamma^n_\mp\}=\dlt^{mn}, m,n=1,2,...,d
   \]
   Now let
   \[
    D_{(k)}=\sum_{k=1}^d{\sum_{s=\pm}{\Gamma^k_s\partial^s_k}}
   \]
   and one can check the ``square root property''
   \[
    (D_{(k)})^2=\sum_{k=1}^d{\partial^+_k \partial^-_k}
   \]
   B. Iochum \etal gave a confirmative answer to the question whether there exists a \dopv
   which gives a desired metric on a finite space in \cite{iochum}. For
   infinite case, as we have mentioned, the sequential works
   \cite{bimonte}\cite{atzmon}\cite{dimakis2} together
   with ours show that this problem is highly
   nontrivial. Besides, work on the implications of our ``natural'' \dopv on
   physics is in proceeding.\\

   {\bf Acknowledgements}\\
    This work was supported by Climb-Up (Pan Deng) Project of
    Department of Science and Technology in China, Chinese
    National Science Foundation and Doctoral Programme Foundation
    of Institution of Higher Education in China.
    One of the authors J.D. is grateful to Dr. Hua Yang in Princeton for pointing out
    one mistake in the proof, to Dr. L-G. Jin in Peking University
    and Dr. H-L. Zhu in Rutgers University
    for their careful reading on this manuscript.
  
 \end{document}